\documentclass[letterpaper, 10 pt, conference]{ieeeconf}  

\IEEEoverridecommandlockouts                              
\usepackage{graphics} 
\usepackage{epsfig} 
\usepackage{mathptmx} 
\usepackage{times} 
\usepackage{amsmath} 
\usepackage{amssymb}  
\usepackage{epstopdf}
\usepackage{subfigure}
\usepackage{epic}
\usepackage{tikz}
\usepackage{algorithm}
\usepackage{algpseudocode}

\usepackage{graphicx} 
\usepackage{graphics} 

\usepackage{color} 
\newtheorem{definition}{Definition}
\newtheorem{theorem}{Theorem}

\newtheorem{remark}{Remark}
\newtheorem{assumption}{Assumption}
\newtheorem{proposition}{Proposition}
\newtheorem{Problem}{Problem}

\floatname{algorithm}{Algorithm}

\title{\LARGE \bf Time-constrained multi-agent task scheduling based on \\prescribed performance control
}

\author{Pian Yu and Dimos V. Dimarogonas
\thanks{This work was supported in part by the Swedish Research Council (VR), the Swedish Foundation for Strategic Research (SSF), the Knut and Alice Wallenberg Foundation (KAW) and the SRA TNG ICT project TOUCHES.}
\thanks{The authors are with School of Electrical Engineering and Computer Science, KTH Royal Institute of Technology, 10044 Stockholm, Sweden.
        {\tt\small piany@kth.se, dimos@kth.se}}
}

\begin{document}

\maketitle
\thispagestyle{empty}
\pagestyle{empty}

\begin{abstract}
The problem of time-constrained multi-agent task scheduling and control synthesis is addressed. We assume the existence of a high level plan which consists of a sequence of cooperative tasks, each of which is associated with a deadline and several Quality-of-Service levels. By taking into account the reward and cost of satisfying each task, a novel scheduling problem is formulated and a path synthesis algorithm is proposed. Based on the obtained plan, a distributed hybrid control law is further designed for each agent. Under the condition that only a subset of the agents are aware of the high level plan, it is shown that the proposed controller guarantees the satisfaction of time constraints for each task. A simulation example is given to verify the theoretical results.
\end{abstract}

\section{INTRODUCTION}

The integration of multi-agent cooperative control and task scheduling is of great practical interest for applications such as robotics. Over the past decades, the research in multi-agent cooperative control has usually focused on achieving one single global task, such as reference-tracking \cite{Ji07}, consensus \cite{Ren05} or formation \cite{Guo14}. In practice, a group of agents encounters the request of a sequence of tasks. Furthermore, deadline constraints on the completion of each task is a common requirement, e.g., ``Visit region A within 10 time units". How to jointly schedule the time-constrained task sequence and design the distributed controllers for the group of agents is a more recent challenge.

Task scheduling is one of the fundamental issues in the area of real-time systems \cite{Ram84}. The scheduling algorithms can be divided into two categories: static and dynamic scheduling. Rate monotonic (RM) scheduling \cite{Liu73} and its extensions \cite{Klein12} are static scheduling algorithms and represent one major paradigm, while many dynamic scheduling algorithms are based on Earliest Deadline First (EDF) policies \cite{Liu73}, which represent a second major paradigm. Under certain conditions, EDF has been shown to be optimal in resource-efficient environments \cite{Stan12}. However, these scheduling algorithms usually do not consider the reward or/and cost of completion each task. Different from the above algorithms, in this paper, a scheduling algorithm is proposed for a sequence of tasks by taking into account the reward and cost of completing each task. Motivated by \cite{Lu02}, the reward is defined based on the Quality-of-Service (QoS) level, which is determined by the completion time. Moreover, the cost is defined as the (estimated) total distance travelled by the group of agents.

Cooperative control of multi-agent systems (MAS) has traditionally focused on designing local control laws to achieve a global control objective. Recently, prescribed performance control (PPC) \cite{Bech08} has been proposed to tackle multi-agent control problems with transient performance constraints. In \cite{Kara12, Luca17}, consensus of MAS with prescribed performance on the position error or combined error was investigated. In \cite{Bech18, Ver18}, formation control of MAS and large vehicular platoons were investigated with prescribed transient and steady-state performance. In \cite{Lars17}, PPC was utilized to satisfy temporal logic tasks for MAS. In our work, to guarantee that each task is completed at certain QoS level, it is required that each task is completed at a specific time interval, for example, ``Visit region A within 6 to 8 time units". The intuition for the use of PPC is that the time interval constraints under consideration can actually be translated into transient performance constraints, and thus PPC can be applied.

Motivated by the above discussion, this paper investigates the problem of time-constrained multi-agent task scheduling and control synthesis. The contributions of the paper can be summarized as: i) a novel scheduling problem is formulated for MAS subject to a sequence of cooperation tasks, where each task is associated with a deadline and several QoS levels; ii) under the condition that only a subset of agents are aware of the high level plan, a distributed hybrid control law is designed for each agent that guarantees the achievement of each task, according to certain time interval constraints.

The rest of the paper is organized as follows. In Section II, notation and preliminaries are introduced, while Section III formalizes the considered problem. Section IV presents the proposed solution in detail, which is further verified by simulations in Section V. Conclusions are given in Section VI.

\section{Preliminaries}

\subsection{Notation}
Let $\mathbb{R}:=(-\infty, \infty)$, $\mathbb{R}_{\ge 0}:=[0, \infty)$, $\mathbb{R}_{> 0}:=(0, \infty)$, $\mathbb{Z}_{> 0}:=\{1,2,\ldots\}$ and $\mathbb{Z}_{\ge 0}:=\{0,1,2,\ldots\}$.  Denote $\mathbb{R}^n$ as the $n$ dimensional real vector space, $\mathbb{R}^{n\times m}$ as the $n\times m$ real matrix space. $I_n$ is the identity matrix of order $n$. For $(x_1, x_2, \ldots, x_m)\in \mathbb{R}^{n_1+n_2+\cdots +n_m}$, the notation $(x_1, x_2, \ldots, x_m)$ stands for $[x_1^T, x_2^T, \ldots, x_m^T]^T$. Let $\left|\lambda\right|$ be the absolute value of a real number $\lambda$, $\|x\|$ and $\|A\|$ be the Euclidean norm of vector $x$ and matrix $A$, respectively. For a set $\Omega$, $|\Omega|$ represents the cardinality of $\Omega$. In addition, we use $\cap$ to denote the set intersection and $\cup$ the set union. $P \succ 0$ means that $P$ is a positive definite matrix and $P^T$ is the transpose of $P$. The Kronecker product is denoted by $\otimes$.

\subsection{Graph Theory}

Let $\mathcal{G}=\{\mathcal {V}, \mathcal {E}\}$ be a graph with the set of nodes $\mathcal {V}={1, 2,\dots, N}$, and $\mathcal{E}\subseteq \{(i,j): i,j\in \mathcal {V},j \ne i\}$ the set of edges. If $(i,j)\in \mathcal {E}$, then node $j$ is called a neighbor of node $i$ and node $j$ can receive information from node $i$. The neighboring set of node $i$ is denoted by $\mathcal{N}_i = \{j \in \mathcal {V} | (j, i)\in \mathcal {E}\}$.

A graph is called undirected if $(i,j)\in \mathcal {E} \Leftrightarrow (j,i)\in \mathcal {E}$, and a graph is connected if for every pair of nodes $(i,j)$, there exists a path which connects $i$ and $j$, where a path is an ordered list of edges such that the head of each edge is equal to the tail of the following edge.
%
%

\section{Problem Formulation}

\subsection{Agent dynamics}

Consider a group of $N$ agents, each of which obeys the second-order dynamics:
\begin{equation}\label{x}
\begin{array}{l}
{\dot x}_i(t) = v_i(t),\\
{\dot v}_i(t) = u_i(t), \quad i = 1,2, \ldots ,N.
\end{array}
\end{equation}
where $x_i\in \mathbb{R}^{n}, v_i\in \mathbb{R}^{n}$ and $u_i\in \mathbb{R}^{n}$ are the position, velocity and control input of agent $i$, respectively. Let $x=(x_1, \ldots, x_N), v=(v_1, \ldots, v_N)$ be the stack vector of positions and velocities, respectively. Denote by $\bar x=(\bar x_1, \ldots, \bar x_m)$ the $m-$dimensional stack vector of relative positions of pairs of agents that form an edge in $\mathcal{G}$, where $m$ is the number of edges. The elements of vector $\bar x$ are defined by $\bar x_k\buildrel \Delta \over = x_{ij}=x_i-x_j$, with $k\in \{1,2,\ldots, m\}$. It is assumed that the graph $\mathcal{G}$ is undirected and connected.

\subsection{Task Specifications}

For the group of agents, we assume the existence of a high level plan $\phi$ given and known by only a subset of the agents that consists of the achievement of a sequence of cooperative tasks. Then, we can denote $\phi$ as a sequence of $M\in \mathbb{Z}_{>0}$ objectives $\phi_1, \phi_2, \ldots, \phi_M$. Note that the order at which the tasks should be completed is not predefined in this paper.

To be more specific, we consider that the group of agents have a set of regions of interest, denoted by $\mathbb{X}\triangleq \{\mathbb{X}_1, \ldots, \mathbb{X}_M\}$, where $\mathbb{X}_l,  l=1, \ldots, M$ is the target region for task $\phi_l$. For simplicity of presentation, $\mathbb{X}_l\in \mathbb{X}$ is represented by a sphere area around a point of interest:
\begin{equation*}
\mathbb{X}_l=\mathcal{B}(c_l, r_l)=\{z\in \mathbb{R}^n: \|z-c_l\|\le r_l\},
\end{equation*}
where $c_l\in \mathbb{R}^n$ is the center, $r_{\rm min}\le r_l\le r_{\rm max}\in \mathbb{R}$ is the radius and $0<r_{\rm min}<r_{\rm max}$ are the upper and lower bounds of the radius for all regions. Define $\mathbb{X}_0:=\{x_1(t_0), \ldots, x_N(t_0)\}$, which represents a collection of the agents' initial states.

\begin{assumption}\label{ass1}
 The initial state set and target sets do not intersect, i.e., $\mathbb{X}_{l_1}\cap \mathbb{X}_{l_2}=\varnothing, \forall l_1, l_2\in \{0, 1, \ldots, M\}, l_1\neq l_2$.
\end{assumption}

Each task has several Quality-of-Service (QoS) levels, which are determined by the (actual) completion time. In this task model, each task $\phi_{l}$ is characterized by the following parameters:

$D_l$: the absolute deadline;

$k_l$: number of QoS levels ($k_l\in \mathbb{Z}, k_l\ge 2$);

$AE_l$: the (actual) completion time;

In addition, for each QoS level $\hat k_l, \hat k_l\in \{0, 1,\ldots, k_l-1\}$, there are two more parameters:

$EE_l[\hat k_l]$: the estimated completion time of task $\phi_l$ at QoS level $\hat k_l$;

$R_l[\hat k_l]:$ the reward that task $\phi_l$ contributes if it is completed at QoS level $\hat k_l$. QoS level 0 represents the rejection of the task and $R_l[0]<0$ is called the rejection penalty \cite{Abde98, Lu02}.

The QoS levels are defined as follows: for each task $\phi_l$, divide the time interval $[t_0, D_l]$ into $k_l-1$ disjoint parts $(t_l^1, D_l], (t_l^2, t_l^1], \ldots, (t_0, t_l^{k_l-2}]$ (not necessarily equal division), such that $D_l:=t_l^0>t_l^1>t_l^2>\cdots >t_l^{k_l-1}:=t_0$ (see Fig.1). We say that the task $\phi_l$ is completed at QoS level 0, if $AE_l>D_l$, and we say that the task $\phi_l$ is completed at QoS level $\hat k_l, \hat k_l\ge 1$, if $AE_l\in (t_l^{\hat k_l}, t_l^{\hat k_l-1}]$. In this paper, it is assumed that the QoS levels and the corresponding time intervals are known to each agent. In addition, without loss of generality, we assume $t_0=0, \forall l$.

\begin{figure}[htp]
\centering
\includegraphics[width=.45\textwidth]{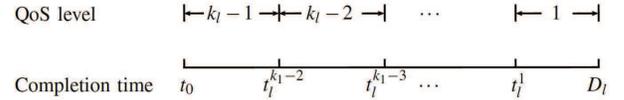}
\caption{Completion time and the corresponding QoS level}
\end{figure}

\subsection{Reward and cost of a path}

In this paper, we are interested in the quantitative reward and cost of satisfying the task plan $\phi$. Let $\Pi_{\phi}$ be the set of permutations of the set $\{1, \ldots, M \}$ and $\pi\in \Pi_{\phi}$ be $\pi:=\{\pi[1], \pi[2], \ldots, \pi[M]\}$. Denote by $\textbf{P}(\pi)=\textbf{P}(\pi[0]\pi[1]\pi[2]\ldots \pi[M])=\mathbb{X}_{\pi[0]}\mathbb{X}_{\pi[1]}\mathbb{X}_{\pi[2]}\cdots\mathbb{X}_{\pi[M]}$, a \emph{path} generated by the task set $\phi$ with the order of completion given by $\pi$, where $\pi[0]=0, \forall \pi\in\Pi_{\phi}$.

1) \emph{Reward}: The reward of a path $\textbf{P}(\pi)$ is given by
\begin{equation}\label{reward}
R(\textbf{P}(\pi))=\sum_{l=1}^M R_{\pi[l]}[\hat k_{\pi[l]}].
\end{equation}

2) \emph{Cost}: Motivated by \cite{Meng16}, the cost of the path $\textbf{P}(\pi)$ is defined as the (estimated) distance travelled by the group of agents. Let $W(\mathbb{X}_{\pi[k]}, \mathbb{X}_{\pi[k+1]}): \mathbb{R}^n \times \mathbb{R}^n \to \mathbb{R}_{\ge 0}$ be the transition cost from set $\mathbb{X}_{\pi[k]}$ to $\mathbb{X}_{\pi[k+1]}$, which is given by
\begin{equation*}
\begin{aligned}
  W(\mathbb{X}_{\pi[k]}, \mathbb{X}_{\pi[k+1]})&\triangleq N(\|c_{\pi[k]}-c_{\pi[k+1]}\|), \; k=1, \ldots, M-1,\\
  W(\mathbb{X}_{\pi[0]}, \mathbb{X}_{\pi[1]})&\triangleq \sum_{i=1}^{N}\|x_i(0)-c_{\pi[1]}\|.
\end{aligned}
\end{equation*}
Then, the cost of $\textbf{P}(\pi)$ is defined as:
\begin{equation}\label{cost}
C(\textbf{P}(\pi))=\sum_{k=0}^{M-1} W(\mathbb{X}_{\pi[k]}, \mathbb{X}_{\pi[k+1]}).
\end{equation}

It is obvious that the completion of the sequence of tasks can be scheduled in different orders, which correspond to different paths. Let $\mathbb{P}=\cup_{\pi\in \Pi} \{\textbf{P}(\pi)\}$ be the set of all paths. In this paper, we want to maximize the reward as well as minimize the cost. Therefore, we propose the following objective function
\begin{equation}\label{objec}
\begin{aligned}
  \textbf{J}(\mathbb{P})\triangleq
  \max_{\pi\in \Pi} \quad \left\{\alpha R(\textbf{P}(\pi))-(1-\alpha)C(\textbf{P}(\pi))\right\},
\end{aligned}
\end{equation}
where $\alpha\in [0,1]$ is a parameter used to balance between the reward and the cost.

Since the high level plan $\phi$ is known to only a subset of the agents, it is necessary for the group of agents to coordinate to complete the tasks. Here, we divide the group of agents into two different groups, active agents (agents who know where the target regions are) and passive agents (agents who do not know where the target regions are). The purpose of this paper is to maximizing the objective function (\ref{objec}) for the set of tasks $\phi$. Formally, the problem is stated below.

\begin{Problem}
Given a group of $N$ agents (which are divided into active agents and passive agents), whose dynamics is given in (\ref{x}), and the task specifications in Section III-B, design distributed (i.e., with only measurements of states of neighbors) control laws $u_i$ and the associated path $\textbf{P}(\pi)$, such that (\ref{objec}) is maximized by $\textbf{P}(\pi)$.
\end{Problem}

\section{Solution}

The proposed solution consists of two layers: i) an offline path synthesis layer, i.e., progressive goal regions and ii) a distributed control law that guarantees that the group of agents (both active and passive) arrive at their progressive goal regions before the corresponding deadline at all times.

\subsection{Path synthesis}

As stated previously, the completion of the sequence of tasks can be scheduled in different order. To find the one that maximizes (\ref{objec}), we propose he following scheduling algorithm $\mathcal{P}(\phi)$:
\begin{subequations}\label{optim}
\begin{eqnarray}
&&\hspace{-1cm}\ \max_{\pi\in \Pi_{\phi}, \hat k_{\pi}} \alpha\sum_{i=1}^{|\pi|}R_{\pi[i]}-(1-\alpha)\sum_{k=1}^{|\pi|-1}W(\mathbb{X}_{\pi[k]}, \mathbb{X}_{\pi[k+1]}) \\
&&\hspace{-1cm}\text{subject to} \nonumber\\
&&\hspace{-0.6cm}\hat k_{\pi[i]}\in \{0, 1, \ldots, k_{\pi[i]}-1\},\label{3c}\\
&&\hspace{-0.6cm}R_{\pi[i]}= R_{\pi[i]}[\hat k_{\pi[i]}], \label{3d}\\
&&\hspace{-0.6cm}EE_{\pi[i]}[0]=D_{\pi[i]}+\epsilon, \label{3e1}\\
&&\hspace{-0.6cm}EE_{\pi[i]}[\hat k_{\pi[i]}]=t_{\pi[i]}^{\hat k_{\pi[i]}-1}, \; \hat k_{\pi[i]}\ge 1\label{3e}\\
&&\hspace{-0.6cm}EE_{\pi[i+1]}[\hat k_{\pi[i+1]}]> EE_{\pi[i]}[\hat k_{\pi[i]}], i=1, \ldots, |\pi|-1. \label{3g}
\end{eqnarray}
\end{subequations}
where
\begin{equation}\label{epsbar}
  \begin{aligned}
  \epsilon=\min\Big\{\Big|t_{l_1}^{\hat k_{l_1}}-t_{l_2}^{\hat k_{l_2}}\Big|: l_1, l_2\in \{1,\ldots, M\}, \hat k_{l_1}\in \{0,\ldots, k_{l_1}-2\}, \\
  \hat k_{l_2}\in \{0,\ldots, k_{l_2}-2\}, l_1\neq l_2, t_{l_1}^{\hat k_{l_1}}\neq t_{l_2}^{\hat k_{l_2}}\Big\}.
  \end{aligned}
\end{equation}
The optimal solution of $\mathcal{P}(\phi)$ is given by $\pi^*=\{\pi^*[1], \ldots, \pi^*[|\mathbb{I}_{\phi}|]\}$ and $\hat k^*_{\pi^*}=\{\hat k^*_{\pi^*[1]}, \dots, \hat k^*_{\pi^*[|\mathbb{I}_{\phi}|]}\}$. In (\ref{3e1}) and (\ref{3e}), the estimated completion time $EE_l[\hat k_l]$ is defined as $D_l+\epsilon$ for QoS level 0 and the deadline of QoS level $\hat k_l$ for $\hat k_l\ge 1$, respectively. Here, $\epsilon$ is a constant used to distinguish between $EE_{\pi[l]}[1]$ and $EE_{\pi[l]}[0]$. Note that tasks with QoS level 0 means the rejection of the task, in other words, the task will not be executed.

\begin{remark}
In general, the algorithm (\ref{optim}) is $NP$ hard since finding the set of permutations is $NP$ hard. However, we note that certain heuristic algorithms such as genetic algorithms \cite{Zach05, Oma10} may be applied in practice.
\end{remark}

\begin{remark}
The obtained plan $\pi^*$ may not be optimal for the objective function (\ref{objec}) because the estimated completion time for each QoS level ($\ge 1$) is given by the deadline of the corresponding time interval (constraint (\ref{3e})). In real-time implementation, a task may be completed before the deadline, which allows for other possibilities of execution. However, we note that the reason for this choice is to guarantee the feasibility of the plan. A way to remedy this is to introduce an online adjustment scheme (e.g., reschedule, restore previous infeasible tasks) at the completion time of each task.
\end{remark}

\subsection{Controller synthesis}

The task execution evolves as follows. Based on the obtained plan $\pi^*$, the task $\pi^*[1]$ will be executed first. Once a task is completed\footnote{A task is said to be completed once all agents lie inside the corresponding target region.}, the agents will proceed immediately to the next one and a new controller will be synthesized and implemented.

Let $\chi$ be a set that keeps track of the executing tasks. The $k$th ($1\le k \le M$) element of $\chi$ is represented by $\chi[k]$, which is given by a triple $\chi[k]=(E_k, t_0^k, T_k)$, and $E_k\in \phi, t_0^k$ and $T_k$ represent the $k$th executing task, the starting time of $E_k$ and the interval of time corresponding to the desired QoS level of $E_k$, respectively. For example, $\chi[2]=(\phi_4, 10, (t_4^2, t_4^1])$ means that the second executing task is $\phi_4$, it is executed from time unit 10, and desired to be completed within the time interval $(t_4^2, t_4^1]$ (QoS level 2). When all tasks are completed, all agents will switch to idle mode, i.e., $E_k=T_k=\varnothing$. Each $\chi[k]$ determines uniquely a control input $u_i$ for each agent $i$, which is denoted by $u_i^{\chi[k]}$.

If $E_k=\varnothing$ (idle mode), the control input is given by
\begin{equation*}
\begin{aligned}
u_i^{\chi[k]} =u_i^{\rm idle}= -\alpha\sum_{j\in \mathcal{N}_i}(x_i-x_j)-\beta v_i,\quad \text{if} \quad E_k=\varnothing,
\end{aligned}
\end{equation*}
where $\alpha, \beta>0$ are positive control gains. Otherwise, we assume $(E_k, t_0^k, T_k)=(\phi_l, t_0^k, (t_l^{\hat k_l}, t_l^{\hat k_l-1}])$. To guarantee that task $\phi_l$ is completed at the desired time interval $(t_l^{\hat k_l}, t_l^{\hat k_l-1}]$, we propose an adaptive controller, which is motivated by prescribed performance control \cite{Bech08}.

\begin{definition}
  A function $\rho: \mathbb{R}_{\ge 0}\to \mathbb{R}_{>0}$ will be called a \emph{performance function} if $\rho$ is continuous, bounded, nonnegative and non-increasing.
\end{definition}

\begin{definition}
  A function $S: \mathbb{R}\to \mathbb{R}$ will be called a \emph{transformation function} if $S$ is strictly increasing, hence injective and admitting an inverse. In particular, let $S_1: [0, 1)\to \mathbb{R}$ with $S_1(z)=\ln(\frac{1}{1-z})$, and $S_2: (-1, 1)\to \mathbb{R}$ with $S_2(z)=\ln(\frac{1+z}{1-z})$.
\end{definition}

Let $\textbf{I}$ be the set of active agents, and $\textbf{F}=\mathcal{V}\setminus {\textbf{I}}$ be the set of passive agents. For active agent $i\in \textbf{I}$, we propose to prescribe the norm of the tracking error $\|x_{i}(t)-c_l\|$ within the following bounds,
\begin{align}\label{ppc1}
  \alpha_{i}^k(t)<\|x_{i}(t)-c_l\|< \beta_{i}^k(t),\quad i\in \textbf{I}.
\end{align}
For passive agent $i\in \textbf{F}$,, we propose to prescribe the norm of the relative distance between neighboring agents within the following bounds,
\begin{align}\label{ppc2}
\|x_{ij}(t)\|< \gamma_i^k(t), \quad i\in\mathcal{V}, (i, j)\in \mathcal{E},
\end{align}
where $\alpha_{i}^k(t), \beta_{i}^k(t), \gamma_i^k(t)$ are performance functions to be defined.

Let $\underline{t}^k=t_l^{\hat k_l}$ and $\bar t^k=t_l^{\hat k_{l-1}}$. The performance functions $\alpha_{i}^k(t), \beta_{i}^k(t), \gamma_i^k(t)$ are non-increasing. Then, to ensure that $AE_l\in (\underline{t}^k, \bar t^k]$, it is sufficient to show that the following two conditions,

C1: $\exists i\in \mathcal{V}$, such that $\|x_i(\underline{t}^k)-c_l\|>r_l$;

C2: $\forall i\in \mathcal{V}$, it holds that $\|x_i(\bar{t}^k)-c_l\|\le r_l$,\\
are satisfied simultaneously.

The scheduling algorithm (\ref{optim}) ensures that $\bar t^k > t_0^k$. However, it is possible that $\underline{t}^k \le t_0^k$. Therefore, in the following, we will present the design of the performance functions and the control synthesis in two different cases, respectively.

\subsubsection{Case I: $\underline{t}^k\le t_0^k$}

Define $\alpha_i^k(t) =0$ and
\begin{subequations}
\begin{eqnarray}
&&\beta_{i}^k(t)=\beta_{i0}^k e^{-\kappa_{i,1}^k (t-t_0^k)}, \quad\quad \label{beta}\\
&&\gamma_i^k(t)=\gamma_{i0}^k e^{-\mu_{i,1}^k (t-t_0^k)}, \quad\quad \label{gamma}
\end{eqnarray}
\end{subequations}
for $t\ge t_0^k$, where $\beta_{i0}^k>\max\{\|x_{i}(t_0^k)-c_l\|, \sigma_k r_l\}$ and $\gamma_{i0}^k>\max\{\max_{j\in \mathcal{N}_i}\{\|x_{ij}(t_0^k)\|\}, r_{\rm min}/(N-1)\}$. In addition,
\begin{subequations}
\begin{eqnarray}
&&\kappa_{i,1}^k=\frac{1}{(\bar t^k-t_0^k)}\ln{\frac{\beta_{i0}^k}{\sigma_k r_l}}, \quad\quad \label{kappa1}\\
&&\mu_{i,1}^k=\frac{1}{(\bar t^k-t_0^k)}\ln{\frac{(N-1)\gamma_{i0}^k}{(1-\sigma_k) r_{\rm min}}}, \quad\quad \label{mu1}
\end{eqnarray}
\end{subequations}
where $\sigma_k\in (0, 1)$.

\begin{remark}\label{rem3}
The definitions of $\beta_{i0}^k, \gamma_{i0}^k$ guarantee that the performance bounds (\ref{ppc1}) and (\ref{ppc2}) are satisfied at starting time $t_0^k$. In addition, from (\ref{beta}) and (\ref{kappa1}), one can get
$$
\|x_i(\bar t^k)-c_l\|<\beta_i^k(\bar t^k)=\beta_{i0}^k e^{-\kappa_{i,1}^k (\bar t^k-t_0^k)}=\sigma_k r_l, \forall i\in \textbf{I}.
$$
Moreover, from (\ref{gamma}) and (\ref{mu1}), one can get \begin{equation*}
\begin{aligned}
\|x_{ij}(\bar t^k)\|<&\gamma_i^k(\bar t^k)=\gamma_{i0}^k e^{-\mu_{i,1}^k (\bar t^k-t_0^k)}\\
=&(1-\sigma_k)r_{\rm min}/(N-1), \forall (i, j)\in \mathcal{E}.
\end{aligned}
\end{equation*}
Since the graph $\mathcal{G}$ is connected, there are at most $N-1$ edges between a passive agent and an active agent, so then one has $\|x_i(\bar t^k)-c_l\|< (N-1)(1-\sigma_k)r_{\rm min}/(N-1)+\sigma_k r_l\le r_l, \forall i\in \textbf{F}$. That is, C2 is satisfied.
\end{remark}

Based on Remark \ref{rem3}, one can conclude that if for task $E_k$, i) the tracking error $\|x_i-c_l\|, \forall i\in \textbf{I}$ is evolving within the prescribed performance bound (\ref{ppc1}), and ii) the relative distance $\|x_{ij}\|, \forall (i,j)\in \mathcal{E}$ is evolving within the prescribed performance bound (\ref{ppc2}) for $t\ge t_0^k$, then the task $E_k$ will be completed within the desired time interval $(\underline t^k, \bar t^k]$.

The approach can be explained as follows. To guarantee that C2 is satisfied, we first ensure that the active agents $i\in \textbf{I}$ will reach the ball area around the target point $c_l$ with radius $\sigma_k r_l$, that is,
\begin{equation}\label{xI}
\mathbb{X}_l^{\textbf{I}}:=\{z\in \mathbb{R}^n: \|z-c_l\|< \sigma_k r_l\},
\end{equation}
before $\bar t^k$. Then, we further ensure that the relative distance between neighboring agents will reach
\begin{equation}\label{xp}
\mathbb{X}_l^{\mathcal{E}}:=\{z\in \mathbb{R}^n: \|z\|< \frac{(1-\sigma_k)r_{\rm{min}}}{N-1}\},
\end{equation}
before $\bar t^k$. As stated in Remark \ref{rem3}, (\ref{xI}) and (\ref{xp}) together imply that C2 is satisfied.

\begin{figure}[htp]
\centering
\includegraphics[width=.3\textwidth]{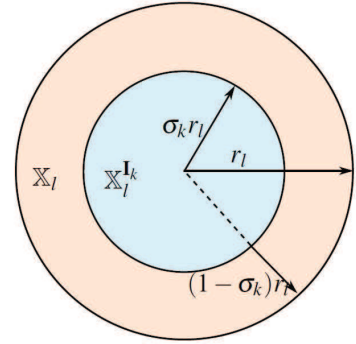}
\caption{Target regions for active agents and passive agents ($n=2$)}
\end{figure}​

\begin{remark}\label{rem5}
We assume that the agents have no information about the communication graph $\mathcal{G}$ (otherwise, a linear feedback controller can be designed, which is similar to \cite{Gui17}). Furthermore, the passive agents do not know if they are connected to active agents. Therefore, in (\ref{xp}), we set the prescribed relative distance between neighboring agents to be $(1-\sigma_k)r_{\rm min}/(N-1)$, which corresponds to the `worst-case' scenario, at the expense of knowing the total number of agents $N$. This level of centralized knowledge can be attained offline and is typical in many multi-agent approaches for less complex problems than the one in hand \cite{Mesb10}.
\end{remark}

Define the normalized errors as
\begin{align}\label{xijk}
  \xi_i(t)=\frac{\|{x_{i}}(t)-c_l\|}{{\beta_{i}^k(t)}},\quad {\xi_{ij}}(t) = \frac{\|{x_{ij}}(t)\|}{{\gamma_i^k(t)}},
\end{align}
respectively. Now, (\ref{ppc1}) is equivalent to $0<{\xi_i}(t)<1$ and (\ref{ppc2}) is equivalent to $0\le{\xi_{ij}}(t)< 1$. The corresponding sets
\begin{equation}\label{D1}
  D_{\xi_i}\triangleq\{{\xi_i}(t): \xi_i(t)\in (0, 1)\}
\end{equation}
\begin{equation}\label{D2}
  D_{\xi _{ij}}\triangleq\{{\xi _{ij}}(t): {\xi _{ij}}(t)\in [0, 1)\}
\end{equation}
are equivalent to (\ref{ppc1}) and (\ref{ppc2}), respectively.

The normalized errors $\xi_i$ and $\xi _{ij}$ are transformed through transformation function $S_1$. We denote the transformed error $\zeta_i(\xi_i)$ and $\varepsilon_{ij}(\xi _{ij})$ by $\zeta_i(\xi_i)=S_1(\xi_i), \varepsilon_{ij}(\xi _{ij}) = S_{1}(\xi _{ij})$, respectively, where we dropped the time argument $t$ for notation convenience.

Let $\nabla S_{1}(\xi_{i})={\partial S_{1}(\xi_{i})}/{\partial \xi_{i}}, \nabla S_{1}(\xi_{ij})={\partial S_{1}(\xi_{ij})}/{\partial \xi_{ij}}$. We propose the following time-varying control protocol:
\begin{equation}\label{u}
\begin{aligned}
u_i^{\chi[k]} =& -\sum_{j\in \mathcal{N}_i}{\frac{1}{\gamma_i^k}\nabla S_{1}(\xi_{ij})\varepsilon_{ij}(\xi _{ij})\textbf{n}_{ij}}\\
&-{\frac{h_i^k}{\beta_{i}^k}\nabla S_{1}(\xi_{i})\zeta_{i}(\xi _{i})\textbf{n}_{i}}-K_{i}^k v_i,
\end{aligned}
\end{equation}
where $h_i^k=1, i\in \textbf{I}$ and $h_i^k=0, i\in \textbf{F}$, $\textbf{n}_{i}=(x_{i}-c_l)/\|x_{i}-c_l\|, \textbf{n}_{ij}=x_{ij}/\|x_{ij}\|$, and $K_{i}^k$ is a positive control gain to be determined later.

Let $\bar \xi_{k} = \xi_{ij}, \bar \varepsilon_{k} = \varepsilon_{ij}$. Let also $\bar\varepsilon=(\bar\varepsilon_1(\bar\xi_1), \ldots, \bar\varepsilon_p(\bar\xi_p)), \bar\zeta=(\zeta_1(\xi_1), \ldots, \zeta_N(\xi_N))$ be the stack vector of the transformed errors. Define $\|z\|':= d\|z\|/dt$. Then, the following holds.

\begin{theorem}\label{thm1}
  Consider the MAS (\ref{x}) and the prescribed performance controller for task $\chi[k]$ given by (\ref{u}) with $\underline{t}^k\le t_0^k$. Suppose Assumption 1 holds and the control gain $K_i^k$ satisfies $K_i^k>\max\{\mu_{i,1}^k, \kappa_{i,1}^k\}, \forall i.$
  Then, i) the tracking error $\|x_i-c_l\|, \forall i\in \textbf{I}$ will evolve within the performance bound (\ref{ppc1}), ii) the relative distance $\|x_{ij}\|, (i,j)\in \mathcal{E}$ will evolve within the performance bound (\ref{ppc2}) for $t\ge t_0^k$, iii) the control signal (\ref{u}) is bounded for a finite completion time.
\end{theorem}

\subsubsection{Case II: $\underline t^k> t_0^k$}

The performance functions $\alpha_{i}^k, \beta_{i}^k, \gamma_i^k$ are defined as
\begin{subequations}
\begin{eqnarray}
&&\hspace{-0.8cm}\alpha_i^k(t) =\Bigg\{\begin{aligned}
&\alpha_{i0}^k e^{-\kappa_{i,2}^k (t-t_0^k)}, \; t\in [t_0^k, \underline t^k]\\
&r_l e^{-\kappa_{i,3}^k (t-\underline t^k)}, \; t>\underline t^k,
\end{aligned}, \label{alpha1}\\
&&\hspace{-0.8cm}\beta_{i}^k(t)=\Bigg\{\begin{aligned}
&\beta_{i0}^k e^{-\kappa_{i,2}^k (t-t_0^k)}, \quad t\in [t_0^k, \underline t^k]\\
&\frac{\beta_{i0}^k r_l}{\alpha_{i0}^k} e^{-\kappa_{i,3}^k (t-\underline t^k)}, \quad t>\underline t^k,
\quad\quad \end{aligned} \label{beta1}\\
&&\hspace{-0.8cm}\gamma_i^k(t)=\Bigg\{\begin{aligned}
&\gamma_{i0}^k e^{-\kappa_{i,2}^k (t-t_0^k)}, \quad t\in [t_0^k, \underline t^k]\\
&\frac{\gamma_{i0}^k r_l}{\alpha_{i0}^k} e^{-\mu_{i,2}^k (t-\underline t^k)}, \quad t>\underline t^k,
\quad\quad \end{aligned} \label{gamma1}
\end{eqnarray}
\end{subequations}​
where
\begin{subequations}
\begin{eqnarray}
&&\alpha_{i0}^k=\|x_{i}(t_0^k)-c_l\|-\Delta_i^k, \label{alpha00}\\
&&\beta_{i0}^k=\|x_{i}(t_0^k)-c_l\|+\Delta_i^k,  \label{beta00}\\
&&\gamma_{i0}^k>\max\Big\{\max_{j\in \mathcal{N}_i}\{\|x_{ij}(t_0^k)\|\}, \frac{r_{\rm min}}{N-1}\Big\},  \label{gamma00}
\end{eqnarray}
\end{subequations}
and $0< \Delta_i^k <\|x_{i}(t_0^k)-c_l\|-r_l, \forall i\in \textbf{I}$. In addition,
\begin{subequations}
\begin{eqnarray}
 &&\hspace{-0.5cm}\kappa_{i,2}^k=\frac{1}{(\underline t^k-t_0^k)}\ln{\frac{\alpha_{i0}^k}{r_l}}, \quad\quad \label{kappa2}\\
 &&\hspace{-0.5cm}\kappa_{i,3}^k=\frac{1}{(\bar t^k-\underline t^k)}\ln{\frac{\beta_{i0}^k}{\sigma_k\alpha_{i0}^k}}, \quad\quad \label{kappa3}\\
 &&\hspace{-0.5cm}\mu_{i,2}^k=\frac{1}{(\bar t^k-\underline t^k)}\ln{\frac{(N-1)\gamma_{i0}^k}{(1-\sigma_k) r_{\rm min}}},\quad\quad \label{mu2}
\end{eqnarray}
\end{subequations}
where $\sigma_k\in (0, 1)$ is a constant to be determined later. One can verify that the functions $\alpha_{i}^k, \beta_{i}^k, \gamma_i^k$ satisfy the definition of performance function.


\begin{proposition}
The performance functions $\alpha_{i}^k, \beta_{i}^k$ and $\gamma_{i}^k$ defined in (\ref{alpha1}), (\ref{beta1}) and (\ref{gamma1}) guarantee that the conditions C1 and C2 are satisfied simultaneously.
\end{proposition}

Let $\rho_{i}^k(t) \buildrel \Delta \over ={(\beta_{i}^k(t)+\alpha_{i}^k(t))}/{2}, \delta_i^k(t) \buildrel \Delta \over = {(\beta_{i}^k(t)-\alpha_{i}^k(t))}/{2}$.
Then, (\ref{ppc1}) can be rewritten as $-\delta_{i}^k(t)+\rho_{i}^k(t)< \|x_{i}(t)-c_l\|< \rho_{i}^k(t)+\delta_{i}^k(t).$ Define in this case the normalized error $\xi_i(t)$ as $\xi_i(t)={({\|{x_{i}}(t)-c_l\|-\rho_{i}^k(t)})}/{{{\delta_{i}^k}(t)}},
$ and $\xi_{ij}$ is defined the same as in (\ref{xijk}). Then, the corresponding set
\begin{equation}\label{D3}
  \hat{D}_{\xi_i}\triangleq\{{\xi_i}(t): \xi_i(t)\in (-1, 1)\}
\end{equation}
is equivalent to (\ref{ppc1}).

The normalized errors $\xi_i$ and $\xi _{ij}$ are transformed through transformation functions $S_2$ and $S_1$, respectively. We denote the transformed errors $\zeta_i(\xi_i)$ and $\varepsilon_{ij}(\xi _{ij})$ by $\zeta_i(\xi_i)=S_2(\xi_i), \varepsilon_{ij}(\xi _{jk}) = S_1(\xi _{ij}).$

Let $\nabla S_{2}(\xi_{i})={\partial S_{2}(\xi_{i})}/{\partial \xi_{i}}, \nabla S_{1}(\xi_{ij})={\partial S_{1}(\xi_{ij})}/{\partial \xi_{ij}}$. We propose the following time-varying control protocol:
\begin{equation}\label{u2}
\begin{aligned}
u_i^{\chi[k]} =& -\sum_{j\in \mathcal{N}_i}{\frac{1}{\gamma_i^k}\nabla S_{1}(\xi_{ij})\varepsilon_{ij}(\xi _{ij})\textbf{n}_{ij}}\\
&-\frac{h_i^k}{\delta_{i}^k}\nabla S_{2}(\xi_{i})\zeta_{i}(\xi _{i})\textbf{n}_{i}-K_{i}^k v_i.
\end{aligned}
\end{equation}
Then, the following holds.

\begin{theorem}\label{thm2}
 Consider the MAS (\ref{x}) and the prescribed performance controller for task $\chi[k]$ given by (\ref{u2}) with $\underline{t}^k> t_0^k$. Suppose Assumption 1 holds and the constants $\sigma_k, K_i^k$ satisfy
  \begin{equation}\label{sigmaK}
  \begin{aligned}
  \sigma_k \le \frac{r_{\rm min}}{(N - 1)\bar\gamma^k+r_{\rm min}},\quad
  K_i^k > \max\left\{\kappa_{i,2}^k, \kappa_{i,3}^k \right\}, \forall i,
  \end{aligned}
  \end{equation}
 where $\bar\gamma^k\ge \max_{i\in \mathcal{V}}\left\{\gamma _{i0}^k\right\}$. Then, i) the tracking error $\|x_i-c_l\|, \forall i\in \textbf{I}$ will evolve within the performance bound (\ref{ppc1}), ii) the relative distance $\|x_{ij}\|, (i,j)\in \mathcal{E}$ will evolve within the performance bound (\ref{ppc2}) for $t\ge t_0^k$, iii) the control signal (\ref{u2}) is bounded for a finite completion time.
\end{theorem}

\begin{remark}
In (\ref{sigmaK}), $\bar\gamma^k$ can be determined as follows. Let $t^*$ be the time instant that algorithm \emph{completion} is activated for the first time. For $t<t^*$, one can choose $\bar\gamma^k=\max\big\{\max_{(i, j)\in \mathcal{E}}\{\|x_{ij}(0)\|\}, {r_{\rm min}}/(N-1)\big\}$. For $t\ge t^*$, one can choose $\bar\gamma^k=2r_{\rm max}$. The task $\phi_l\in\phi$ is completed when all agents lie inside the region $\mathbb{X}_l$. Therefore, at the completion time, one must have $\|x_i-x_j\|\le \|x_i-r_l\|+\|x_j-r_l\|\le 2r_l\le 2r_{\rm max}, \forall (i,j)\in \mathcal{E}, \forall l$. This choice of $\bar\gamma^k$ guarantees that $\bar\gamma^k\ge \max_{i\in \mathcal{V}}\left\{\gamma _{i0}^k\right\}$ is satisfied for all $k$. Another option is $\bar\gamma^k=\max\big\{\max_{(i, j)\in \mathcal{E}}\{\|x_{ij}(0)\|\}, 2{r_{\rm max}}\big\}, \forall k$.
\end{remark}

\begin{remark}
We note that if for task $\phi_l$, one has $\textbf{I}_l=\mathcal{V}$, i.e., all agents are active, then it is possible to design a linear feedback controller for each agent $i$, such that the task $\phi_l$ is completed at any desired QoS level.
\end{remark}

\begin{remark}
In practice, it may not be possible to use prescribed
performance controller to drive an agent to a given target region within any time interval due to physical constraints (e.g., input constraints). However, given the initial conditions and the upper bound of the prescribed performance controller, one may able to calculate (or estimate) the minimal time required to drive one agent to a given target region (e.g., using reachability analysis). Then, the constraint (\ref{3g}) in (\ref{optim}) can be revised as
$$
EE_{\pi[i+1]}[\hat k_{\pi[i+1]}]\ge EE_{\pi[i]}[\hat k_{\pi[i]}]+t_{\rm min}(\mathbb{X}_{\pi[i]}, \mathbb{X}_{\pi[i+1]}),
$$
where $t_{\rm min}(\mathbb{X}_{\pi[i]}, \mathbb{X}_{\pi[i+1]})$ represents the minimal time required to navigate from region $\mathbb{X}_{\pi[i]}$ to $\mathbb{X}_{\pi[i+1]}$.
\end{remark}

\section{Simulation}

In this section, a numerical example is given to verify the theoretical results. Consider a network of 4 agents with $n=2$, and the communication graph $\mathcal{G}$ is shown in Fig. 3, where agent 1 is active and agents 2,3,4 are passive. The initial position $x_i(0)$ of each agent $i$ is chosen randomly from the box $[0,2]\times [0,2]$, and the initial velocity $v_i(0)$ of each agent $i$ is chosen to be $[0,0]^T, \forall i$.

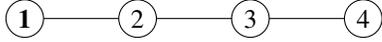
\begin{figure}[htp]
\centering
\begin{tikzpicture}[scale=0.5]
\setlength{\unitlength}{2mm}
\draw (2,.5) circle (.5cm);
\draw (5,.5) circle (.5cm);
\draw (8,.5) circle (.5cm);
\draw (11,.5) circle (.5cm);
\draw[-] (2.5,.5) -- (4.5,.5);
\draw[-] (5.5,.5) -- (7.5,.5);
\draw[-] (8.5,.5) -- (10.5,.5);
\draw (2,.5) node {\textbf{1}};
\draw (5,.5) node {2};
\draw (8,.5) node {3};
\draw (11,.5) node {4};
\end{tikzpicture}
\caption{Communication graph among the agents.}
\end{figure}

The high level plan $\phi$ consists of 3 tasks $\phi_1, \phi_2, \phi_3$, the corresponding target sets are given by:
$\mathbb{X}_1=\mathcal{B}([10, 8]^T, 1)$ with deadline $D_1=15$; $\mathbb{X}_2=\mathcal{B}([3, 10]^T, 1)$ with deadline $D_2=20$; $\mathbb{X}_3=\mathcal{B}([5, 5]^T, 1)$ with deadline $D_3=24$. Tasks $\phi_1, \phi_3$ have 4 QoS levels, while task $\phi_2$ has 2 QoS levels, respectively. The corresponding time intervals and rewards are given by
\begin{equation*}
\begin{aligned}
\phi_1:& \{(15,+\infty), -20\}, \{(10,15], 5\}, \{(5,10], 10\}, \{(0,5], 8\};\\
\phi_2:& \{(20,+\infty), -20\}, \{(0,20], 10\};\\
\phi_3:& \{(24,+\infty), -20\}, \{(14,24], 5\}, \{(9,14], 10\}, \{(0,9], 5\}.
\end{aligned}
\end{equation*}
The objective function is given in (\ref{objec}) with $\alpha=0.8$. Then by solving (\ref{optim}), one can get that the optimal solution is $\{(\pi^*[1], \hat k^*_{\pi^*[1]}), (\pi^*[2], \hat k^*_{\pi^*[2]}), (\pi^*[3], \hat k^*_{\pi^*[3]})\}=\{(1, 3), (3, 3), (2, 2)\}$.

The simulation results are shown in Figs. 4-6. Fig.4 shows the evolution of positions for each agent, where $x_{i1}$ and $x_{i2}$ are position components. The three red circles (from left to right) represent the three target regions $\mathbb{X}_1, \mathbb{X}_3$ and $\mathbb{X}_2$, respectively. The evolution of tracking error $\|x_1-c_l\|$ for active agent 1 and the performance bounds $\alpha_1^k, \beta_1^k$ are depicted in Fig. 5. In addition, the evolution of relative distances between neighboring agents and the performance bounds $\gamma_{1,2,3,4}^k$ are plotted in Fig. 6. One can see that the performance bounds are satisfied at all times.

\begin{figure}[thpb]
\centering
\includegraphics[height=4.5cm,width=6cm]{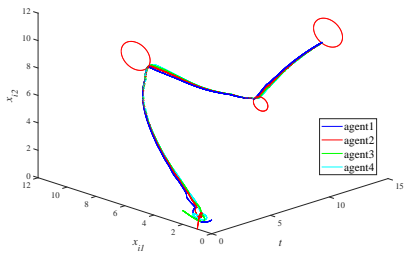}
\caption{The evolution of positions for each agent under (\ref{u}).}
\label{fig5}
\end{figure}

\begin{figure}[thpb]
\centering
\includegraphics[height=4cm,width=8cm]{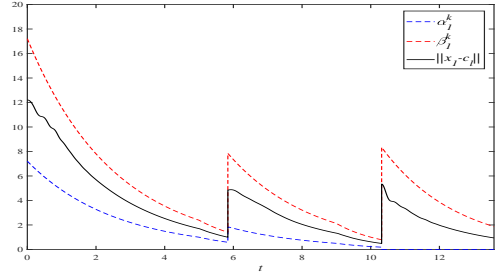}
\caption{The evolution of the tracking error for active agent 1 and performance bounds. }
\end{figure}

\begin{figure}[thpb]
\centering
\includegraphics[height=4cm,width=8cm]{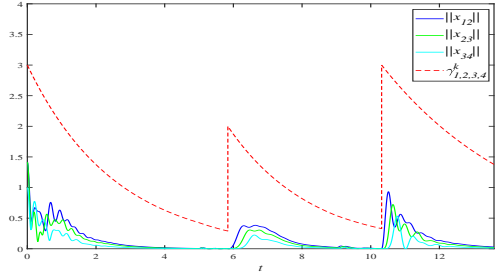}
\caption{The evolution of the relative distances between neighboring agents and performance bounds.}
\end{figure}


\section{Conclusions}

We proposed a task scheduling algorithm and distributed control design for the coordination of MAS that is requested to visit a sequence of target regions with deadline constraints. By utilizing ideas from prescribed performance control, we developed a hybrid feedback control law that guarantees the satisfaction of each task under specific time interval constraints. A natural next step is to consider more complex task specifications and perform physical experiments.


\section*{Appendix}

\emph{Proof of Theorem 1}: Let $y_i=x_i-c_l$ and $y=(y_1, \ldots, y_N)$. Consider the following function
\begin{equation}\label{V}
\begin{aligned}
V_1(y, v,\bar\varepsilon, \bar\zeta) = &\frac{1}{2}[y \quad v]\Bigg\{\left[ \begin{array}{l}
K^k\theta^k \quad \theta^k\\
\quad \theta^k\quad {I_{N}}
\end{array} \right]\otimes I_n\Bigg\}\left[ \begin{array}{l}
y\\
v
\end{array} \right] \\
&+ \frac{1}{2}\bar\varepsilon^T\bar\varepsilon+ \frac{1}{2}\bar\zeta^T H^k \bar\zeta,
\end{aligned}
\end{equation}
where $H^k\in \mathbb{R}^{N\times N}$ is a diagonal matrix with entries $h_i^k$, $K^k\in \mathbb{R}^{N\times N}$ is a diagonal matrix with entries $K_i^k$, and $\theta^k$ is a diagonal matrix with entries $\theta_i^k=\max\{\mu_{i,1}^k, \kappa_{i,1}^k\}$. Since $K_i^k>\max\{\mu_{i,1}^k, \kappa_{i,1}^k\}=\theta_i^k, \forall i$, one can derive that $V(y, v,\bar\varepsilon, \bar\zeta)$ is positive definite for all $t\ge t_0^k$.

Differentiating (\ref{V}) along the trajectories of (\ref{x}), one has
\begin{equation}\label{V1}
\begin{aligned}
\dot V_1(y, v,\bar\varepsilon, \bar\zeta) = & y^T K^k\theta^k v+
y^T\theta^k u^{\chi[k]}+ v^T\theta^k v+v^Tu^{\chi[k]}\\
&+\bar\varepsilon^T\dot{\bar\varepsilon}+\bar\zeta^T H^k \dot{\bar\zeta}.
\end{aligned}
\end{equation}
Substituting (\ref{u}) into (\ref{V1}), we obtain
\begin{equation}\label{V2}
\begin{aligned}
\dot V_1(y, v,\bar\varepsilon, \bar\zeta) =
&-\sum\limits_{i = 1}^N {\theta_i^k y_i^T\sum\limits_{j \in {\mathcal{N}_i}}\frac{1}{\gamma_i^k}\nabla S_{1}(\xi_{ij}){\varepsilon_{ij}(\xi _{ij})}}{ \textbf{n}_{ij}} \\
 &-\sum\limits_{i = 1}^N \theta_i^k{y_i^T}\frac{h_i^k}{\beta_i^k} \nabla S_{1}(\xi_{i})\zeta_{i}(\xi _{i})\textbf{n}_{i} \\
 &- \sum\limits_{i = 1}^N {v_i^T\sum\limits_{j \in {\mathcal{N}_i}}} \frac{1}{\gamma_i^k}\nabla S_{1}(\xi_{ij}){\varepsilon_{ij}(\xi _{ij})}{\textbf{n}_{ij}}\\
 &- \sum\limits_{i = 1}^N {v_i^T} \frac{h_i^k}{\beta_i^k}\nabla S_{1}(\xi_{i})\zeta_{i}(\xi _{i})\textbf{n}_{i}\\
 &-\sum_{i=1}^N(K_i^k-\theta_i^k ){v_i^T}v_i+\sum\limits_{i=1}^p {{\varepsilon _{ij}}({\xi _{ij}})\nabla S_{1}(\xi_{ij})} \dot\xi _{ij}\\
 &+\sum\limits_{i = 1}^N h_i^k{{\zeta _i}} ({\xi _i})\nabla {S_1}({\xi _i})\dot\xi_i.
\end{aligned}
\end{equation}
According to (\ref{xijk}), one can get
\begin{equation*}
\begin{aligned}
\dot\xi _{ij}=&{\frac{1}{\gamma_i^k}\frac{ \|x_{ij}\|'\gamma_i^k-\|x_{ij}\|\dot\gamma_i^k }{\gamma_i^k}}={\frac{1}{\gamma_i^k}\Big( \|x_{ij}\|'+ \mu_{i,1}^k \|x_{ij}\| \Big)},\\
\dot\xi_i=&{\frac{1}{\beta_i^k}\frac{\|x_i-c_l\|'\beta_i^k- \|x_i-c_l\|\dot\beta_i^k}{\beta_i^k}} \\
=& \frac{1}{\beta_i^k}\Big( \|x_i-c_l\|'+ \kappa_{i,1}^k\|x_i-c_l\| \Big)\\
=& \frac{1}{\beta_i^k}\Big( \|y_i\|'+ \kappa_{i,1}^k\|y_i\| \Big),
\end{aligned}
\end{equation*}
where $-\dot\beta_i^k/\beta_i^k\equiv \kappa_{i,1}^k$ and $-\dot\gamma_i^k/\gamma_i^k \equiv \mu_{i,1}^k$.

Due to symmetry, one has
\begin{equation*}
  \frac{{\partial \left\| {{x_{ij}}} \right\|}}{{\partial {x_{ij}}}} = \frac{{\partial \left\| {{x_{ij}}} \right\|}}{{\partial {x_i}}} =  - \frac{{\partial \left\| {{x_{ij}}} \right\|}}{{\partial {x_j}}},
\end{equation*}
and from (\ref{x}),
\begin{equation*}
\begin{aligned}
&\sum\limits_{i=1}^n{\nabla S_{1}(\xi_{ij}){\varepsilon _{ij}}({\xi _{ij}}){{\left\| {{x_{ij}}} \right\|}'}}  \\
=& \frac{1}{2}\sum\limits_{i=1}^N \sum\limits_{j\in \mathcal{N}_i} {\nabla S_{1}(\xi_{ij}){\varepsilon _{ij}}({\xi _{ij}})\frac{{\partial \left\| {{x_{ij}}} \right\|}}{{\partial {x_{ij}}}}{{\dot x}_{ij}}} \\
=& \sum\limits_{i = 1}^N {v_i^T} \sum\limits_{j\in \mathcal{N}_i} {\nabla S_{1}(\xi_{ij}){\varepsilon _{ij}}({\xi _{ij}})}\textbf{n}_{ij}.
\end{aligned}
\end{equation*}
In addition,
\begin{equation*}
\begin{aligned}
&\sum\limits_{i = 1}^N {y_i^T\sum\limits_{j \in {\mathcal{N}_i}} {\nabla S_{1}(\xi_{ij}){\varepsilon_{ij}(\xi _{ij})}} } {\textbf{n}_{ij}}\\
=& \sum\limits_{i = 1}^N {y_i^T\sum\limits_{j \in {\mathcal{N}_i}} {\nabla S_{1}(\xi_{ij}){\varepsilon_{ij}(\xi _{ij})}} } {\frac{y_i-y_j}{\|y_i-y_j\|}}\\
=&\frac{1}{2}\sum\limits_{i = 1}^N {\sum\limits_{j \in {\mathcal{N}_i}} {\nabla S_{1}(\xi_{ij}){\varepsilon_{ij}(\xi _{ij})}} } \left\| {{y_{ij}}} \right\|\\
=&\frac{1}{2}\sum\limits_{i = 1}^N {\sum\limits_{j \in {\mathcal{N}_i}} {\nabla S_{1}(\xi_{ij}){\varepsilon_{ij}(\xi _{ij})}} } \left\| {{x_{ij}}} \right\|.
\end{aligned}
\end{equation*}
Then, (\ref{V2}) can be rewritten as
\begin{equation}\label{V3}
\begin{aligned}
\dot V_1(y, v,\bar\varepsilon, \bar\zeta) =
&-\frac{1}{2}\sum\limits_{i=1}^N \sum\limits_{j\in \mathcal{N}_i}{\frac{\theta_i^k-\mu_{i,1}^k}{\gamma_i^k}\nabla S_{1}(\xi_{ij}){\varepsilon _{ij}}({\xi _{ij}})}\|x_{ij}\| \\
&-\sum\limits_{i = 1}^N {\frac{h_i^k(\theta_i^k-\kappa_{i,1}^k)}{\beta_i^k}{\zeta _i}} ({\xi _i})\nabla {S_1}({\xi _i})\|y_i\| \\
 &-(K^k -\theta^k ){v^T}v,\\
\end{aligned}
\end{equation}
According to the definition of $S_1$, one can derive that $\nabla S_{1}(\xi_{ij}){\varepsilon _{ij}}({\xi _{ij}})\|x_{ij}\|\ge 0$ and ${\zeta _i}({\xi _i})\nabla {S_1}({\xi _i})\|y_i\|\ge 0$. In addition, $\theta_i^k-\mu_{i,1}^k\ge 0$ and $\theta_i^k-\kappa_{i,1}^k\ge 0$ for all $i$. Therefore, one derives that $\dot V_1(y, v,\bar\varepsilon, \bar\zeta)\le 0,$ which in turn implies $V_1(y, v,\bar\varepsilon, \bar\zeta)\le V_1(y(t_0^k), v(t_0^k),\bar\varepsilon(t_0^k), \bar\zeta(t_0^k):= V_1(t_0^k)$ and thus
$$
|\zeta_i(\xi_i)|\le |\bar\zeta|\le \sqrt{2V_1(t_0^k)}, \forall i\in \textbf{I}_k
$$
and
$$
|\varepsilon_{ij}(\xi_{ij})|\le |\bar \varepsilon|\le \sqrt{2V_1(t_0^k)}, \forall (i,j)\in \mathcal{E},
$$
for all $t\ge t_0^k$. Moreover, $\xi_i(t_0^k), \forall i\in \textbf{I}_k$ and $\xi_{ij}(t_0^k), \forall(i,j)\in \mathcal{E}$ are within the regions (\ref{D1}) and (\ref{D2}), respectively. By using the inverse of $S_1$, we can bound $0\le\xi_{i}(t)\le S_{1}^{-1}\Big(\sqrt{2V_1(t_0^k)}\Big)<1$ and $0\le\xi_{ij}(t)\le S_{1}^{-1}\Big(\sqrt{2V_1(t_0^k)}\Big)<1$ for all $t>t_0^k$. That is to say, $\xi_i(t), \forall i\in \textbf{I}_k$ and $\xi_{ij}(t), \forall(i, j)\in \mathcal{E}$ will evolve within the regions (\ref{D1}) and (\ref{D2}) for all $t\ge t_0^k$.

Since $\xi_i(t), \xi_{ij}(t)\in \Big[0, S_1^{-1}\Big(\sqrt{2V_1(t_0^k)}\Big)\Big)$, one has that $\nabla S_{1}(\xi_{i})$ is bounded for all $i\in \textbf{I}_k$ and $\nabla S_{1}(\xi_{ij})$ is bounded for all $(i, j)\in \mathcal{E}$. Furthermore, $\gamma_i^k, \beta_i^k$ are continuous and $0<\gamma_i^k<\infty, 0<\beta_i^k<\infty$ for a finite completion time. Therefore, one can conclude that the control signal (\ref{u}) is bounded for a finite completion time. $\square$

\emph{Proof of Theorem 2}: Let $z=(y, v)^T$. Consider the following function
\begin{equation}\label{V4}
V_2(z,\bar\varepsilon, \bar\zeta) =\left\{\begin{aligned}&\frac{1}{2}(z^TG_1z+
+ \bar\varepsilon^T\bar\varepsilon+\bar\zeta^T H^k\bar\zeta),\; t\in [t_0^k, \underline t^k]\\
&\frac{1}{2}(z^TG_2z+
+ \bar\varepsilon^T\bar\varepsilon+\bar\zeta^T H^k\bar\zeta),\; t>\underline t^k\\
\end{aligned}\right.
\end{equation}
where
\begin{equation}\label{G12}
{G_1} = \left( \begin{array}{l}
{K^k}\kappa _2^k \quad \kappa _2^k\\
\;\kappa _2^k \quad \; {I_N}
\end{array} \right) \otimes {I_n}, {G_2} = \left( \begin{array}{l}
{K^k}\kappa _3^k \quad \kappa _3^k\\
\; \kappa _3^k \quad \; {I_N}
\end{array} \right) \otimes {I_n},
\end{equation}
and $\kappa_2^k, \kappa_3^k\in \mathbb{R}^{N\times N}$ are diagonal matrices with entries $\kappa_{i,2}^k$ and $\kappa_{i,3}^k$, respectively. The matrices $H^k, K^k$ are defined as in the proof of Theorem 1.
Since $K_i^k>\max\{\kappa_{i,2}^k, \kappa_{i,3}^k\}, \forall i$, one can derive $G_1\succ 0, G_2\succ 0$. Therefore, $V_2(z,\bar\varepsilon, \bar\zeta)$ is positive definite for all $t\ge t_0^k$.

i) For $t\in [t_0^k, \underline t^k]$, differentiating (\ref{V4}) along the trajectories of (\ref{x}) and substituting (\ref{u2}), one has
\begin{equation}\label{V5}
\begin{aligned}
\dot V_2(z,\bar\varepsilon, \bar\zeta)=&-\sum\limits_{i = 1}^N {\kappa_{i,2}^k y_i^T\sum\limits_{j \in {\mathcal{N}_i}}\frac{1}{\gamma_i^k}\nabla S_{1}(\xi_{ij}){\varepsilon_{ij}(\xi _{ij})}}{ \textbf{n}_{ij}} \\
 &-\sum\limits_{i = 1}^N \kappa_{i,2}^k{y_i^T}\frac{h_i^k}{\delta_i^k} \nabla S_{2}(\xi_{i})\zeta_{i}(\xi _{i})\textbf{n}_{i} \\
 &- \sum\limits_{i = 1}^N {v_i^T\sum\limits_{j \in {\mathcal{N}_i}}} \frac{1}{\gamma_i^k}\nabla S_{1}(\xi_{ij}){\varepsilon_{ij}(\xi _{ij})}{\textbf{n}_{ij}}\\
 &- \sum\limits_{i = 1}^N {v_i^T} \frac{h_i^k}{\delta_i^k}\nabla S_{2}(\xi_{i})\zeta_{i}(\xi _{i})\textbf{n}_{i}\\
 &-\sum_{i=1}^N(K_i^k-\kappa_{i,2}^k ){v_i^T}v_i+\sum\limits_{i=1}^p {{\varepsilon _{ij}}({\xi _{ij}})\nabla S_{1}(\xi_{ij})} \dot\xi _{ij}\\
 &+\sum\limits_{i = 1}^n h_i^k{{\zeta _i}} ({\xi _i})\nabla {S_2}({\xi _i})\dot\xi_i,
\end{aligned}
\end{equation}
where
\begin{equation*}
\begin{aligned}
\dot\xi _{ij}=&{\frac{1}{\gamma_i^k}\frac{ \|x_{ij}\|'\gamma_i^k-\|x_{ij}\|\dot\gamma_i^k }{\gamma_i^k}},\\
\dot\xi_i=&{\frac{1}{\delta_i^k}\frac{(\|x_i-c_l\|'-\dot\rho_i^k)\delta_i^k- (\|x_i-c_l\|-\rho_i^k)\dot\delta_i^k}{\delta_i^k}} \\
=&{\frac{1}{\delta_i^k}\frac{(\|y_i\|'-\dot\rho_i^k)\delta_i^k- (\|y_i\|-\rho_i^k)\dot\delta_i^k}{\delta_i^k}}.
\end{aligned}
\end{equation*}
Similar to the proof of Theorem \ref{thm1}, one can further get
\begin{equation}\label{V6}
\begin{aligned}
\dot V_2(z,\bar\varepsilon, \bar\zeta) =
&-\frac{1}{2}\sum\limits_{i=1}^N \sum\limits_{j\in \mathcal{N}_i}{\frac{\kappa_{i,2}^k-\hat\gamma_i^k}{\gamma_i^k}\nabla S_{1}(\xi_{ij}){\varepsilon _{ij}}({\xi _{ij}})}\|x_{ij}\| \\
&-\sum\limits_{i = 1}^N \kappa_{i,2}^k\frac{h_i^k}{\delta_i^k} \nabla S_{2}(\xi_{i})\zeta_{i}(\xi _{i})\|y_i\| \\
&-\sum\limits_{i = 1}^n \frac{h_i^k}{\delta_i^k}{{\zeta _i}} ({\xi _i})\nabla {S_2}({\xi _i})\dot\rho_i^k\\
&+\sum\limits_{i = 1}^N {\frac{h_i^k\hat\delta_i^k}{\delta_i^k}{\zeta _i}} ({\xi _i})\nabla {S_2}({\xi _i})(\|y_i\|-\rho_i^k) \\
 &-\sum_{i=1}^N(K_i^k -\kappa_{i,2}^k){v_i^T}v_i,\\
 =&-\frac{1}{2}\sum\limits_{i=1}^N \sum\limits_{j\in \mathcal{N}_i}{\frac{\kappa_{i,2}^k-\hat\gamma_i^k}{\gamma_i^k}\nabla S_{1}(\xi_{ij}){\varepsilon _{ij}}({\xi _{ij}})}\|x_{ij}\| \\
&-\sum\limits_{i = 1}^N {\frac{h_i^k(\kappa_{i,2}^k-\hat\delta_i^k)}{\delta_i^k}{\zeta _i}} ({\xi _i})\nabla {S_2}({\xi _i})(\|y_i\|-\rho_i^k) \\
 &-\sum\limits_{i = 1}^N {\frac{h_i^k\rho_i^k(\kappa_{i,2}^k-\hat\rho_i^k)}{\delta_i^k}{\zeta _i}} ({\xi _i})\nabla {S_2}({\xi _i}) \\
 &-\sum_{i=1}^N(K_i^k -\kappa_{i,2}^k){v_i^T}v_i,\\
\end{aligned}
\end{equation}
where $\hat{\gamma}_i^k=-\dot{\gamma}_i^k/\gamma_i^k$, $\hat{\delta}_i^k=-\dot{\delta}_i^k/\delta_i^k$ and $\hat{\rho}_i^k=-\dot{\rho}_i^k/\rho_i^k$. In addition, according to the definition of $\gamma_i^k(t), \delta_i^k(t)$ and $\rho_i^k(t)$, one can derive that $\hat{\gamma}_i^k(t)=\hat{\delta}_i^k(t)=\hat{\rho}_i^k(t)\equiv \kappa_{i,2}^k$ for all $t\in [t_0^k, \underline t^k]$. Therefore,
\begin{equation}\label{V6}
\begin{aligned}
\dot V_2(z,\bar\varepsilon, \bar\zeta)\le
 &-\sum_{i=1}^N(K_i^k -\kappa_{i,2}^k){v_i^T}v_i\le 0, \quad \forall [t_0^k, \underline t^k].
\end{aligned}
\end{equation}

ii) For $t>\underline t^k$, differentiating (\ref{V4}) along the trajectories of (\ref{x}) and substituting (\ref{u2}), one has
\begin{equation}\label{V7}
\begin{aligned}
\dot V_2(z,\bar\varepsilon, \bar\zeta) =
&-\frac{1}{2}\sum\limits_{i=1}^N \sum\limits_{j\in \mathcal{N}_i}{\frac{\kappa_{i,3}^k-\hat\gamma_i^k}{\gamma_i^k}\nabla S_{1}(\xi_{ij}){\varepsilon _{ij}}({\xi _{ij}})}\|x_{ij}\| \\
&-\sum\limits_{i = 1}^N {\frac{h_i^k(\kappa_{i,3}^k-\hat\delta_i^k)}{\delta_i^k}{\zeta _i}} ({\xi _i})\nabla {S_2}({\xi _i})(\|y_i\|-\rho_i^k) \\
 &-\sum\limits_{i = 1}^N {\frac{h_i^k\rho_i^k(\kappa_{i,3}^k-\hat\rho_i^k)}{\delta_i^k}{\zeta _i}} ({\xi _i})\nabla {S_2}({\xi _i}) \\
 &-\sum_{i=1}^N(K_i^k -\kappa_{i,3}^k){v_i^T}v_i,\\
\end{aligned}
\end{equation}
where $\hat{\delta}_i^k(t)=\hat{\rho}_i^k(t)\equiv \kappa_{i,3}^k$ for all $t>\underline t^k$. Then, one can further have
\begin{equation}\label{V8}
\begin{aligned}
\dot V_2(z,\bar\varepsilon, \bar\zeta) =
&-\frac{1}{2}\sum\limits_{i=1}^N \sum\limits_{j\in \mathcal{N}_i}{\frac{\kappa_{i,3}^k-\hat\gamma_i^k}{\gamma_i^k}\nabla S_{1}(\xi_{ij}){\varepsilon _{ij}}({\xi _{ij}})}\|x_{ij}\| \\
&-\sum_{i=1}^N(K_i^k -\kappa_{i,3}^k){v_i^T}v_i.
\end{aligned}
\end{equation}

If the constant $\sigma_k$ satisfies (\ref{sigmaK}), one can verify that $\hat\gamma_i^k<\kappa_{i,3}^k, \forall i$. In addition, one has $\nabla S_{1}(\xi_{ij}){\varepsilon _{ij}}({\xi _{ij}})\|x_{ij}\|\ge 0, \forall (i, j)\in \mathcal{E}$. Therefore,
\begin{equation}\label{V8}
  \dot V_2(z,\bar\varepsilon, \bar\zeta)\le 0, \quad \forall t>\underline t^k.
\end{equation}
Combining (\ref{V6}) and (\ref{V8}), one can get that $V_2(z,\bar\varepsilon, \bar\zeta)\le \max\{V_2(z(t_0^k), \bar\varepsilon(t_0^k), \bar\zeta(t_0^k)), V_2(z(\underline t^k), \bar\varepsilon(\underline t^k), \bar\zeta(\underline t^k))\}:= V_2^*$ and thus
$$
|\zeta_i(\xi_i)|\le |\bar\zeta|\le \sqrt{2V_2^*}, \forall i\in \textbf{I}_k
$$
and
$$
|\varepsilon_{ij}(\xi_{ij})|\le |\bar \varepsilon|\le \sqrt{2V_2^*}, \forall (i,j)\in \mathcal{E},
$$
for all $t\ge t_0^k$. Moreover, $\xi_i(t_0^k), \forall i\in \textbf{I}_k$ and $\xi_{ij}(t_0^k), \forall(i,j)\in \mathcal{E}$ are within the regions (\ref{D3}) and (\ref{D2}), respectively. By using the inverse of $S_1$ and $S_{2}$, we can bound $-1<S_2^{-1}\Big(-\sqrt{2V_2^*}\Big)\le\xi_i(t)\le S_2^{-1}\Big(\sqrt{2V_2^*}\Big)<1, \forall i\in \textbf{I}_k$ and $0\le\xi_{ij}(t)\le S_{1}^{-1}\Big(\sqrt{2V_2^*}\Big)<1, \forall(i,j)\in \mathcal{E}$ for $t>t_0^k$. That is to say, $\xi_i(t), \forall i\in \textbf{I}_k$ and $\xi_{ij}(t), \forall(i, j)\in \mathcal{E}$ will evolve within the regions (\ref{D3}) and (\ref{D2}) for all $t\ge t_0^k$.

The remainder of the proof is similar to that of Theorem \ref{thm1} and hence omitted.  $\square$


\begin{thebibliography}{0}

\bibitem{Ji07}
M. Ji and M. B. Egerstedt, ``Distributed coordination control of multiagent systems while preserving connectedness", \emph{IEEE Trans. Robot.}, Vol. 23, pp. 693-703, 2007.

\bibitem{Ren05}
W. Ren and R. W. Beard, ``Consensus seeking in multiagent systems under dynamically changing interaction topologies", \emph{IEEE Trans.
Autom. Control}, Vol. 50, pp. 655-661, 2005.

\bibitem{Guo14}
M. Guo, M. M. Zavlanos and D. V. Dimarogonas, ``Controlling the relative
agent motion in multi-agent formation stabilization", \emph{IEEE Trans.
Autom. Control}, Vol. 59, pp. 820-826, 2014.

\bibitem{Ram84}
K. Ramamritham and J. A. Stankovic, ``Dynamic task scheduling in hard real-time distributed systems", \emph{IEEE software}, Vol. 1, pp. 65-75, 1984.

\bibitem{Liu73}
C. L. Liu and James W. Layland, ``Scheduling algorithms for multiprogramming in a hard-real-time environment", \emph{Journal of the ACM (JACM)}, Vol. 20, pp. 46-61, 1973.

\bibitem{Klein12}
M. Klein, T. Ralya, B. Pollak, R. Obenza and M. G. Harbour, A practitioner's handbook for real-time analysis: guide to rate monotonic analysis for real-time systems, Springer Science \& Business Media, 2012.

\bibitem{Stan12}
J. A. Stankovic, M. Spuri, K. Ramamritham and G. C. Buttazzo. Deadline scheduling for real-time systems: EDF and related algorithms, Springer Science \& Business Media, 2012.

\bibitem{Bech08}
C. P. Bechlioulis and G. A. Rovithakis, ``Robust adaptive control of feedback linearizable MIMO nonlinear systems with prescribed performance", \emph{IEEE Trans.
Autom. Control}, Vol. 53, pp. 2090-2099, 2008.

\bibitem{Kara12}
Y. Karayiannidis, D. V. Dimarogonas and D. Kragic, ``Multi-agent average consensus control with prescribed performance guarantees", \emph{in Proc. Decision and Control (CDC)}, pp. 2219-2225, 2012.

\bibitem{Luca17}
L. Macellari, Y. Karayiannidis, D. V. Dimarogonas, ``Multi-agent second order average consensus with prescribed transient behavior", \emph{IEEE Trans. Autom. Control}, Vol. 62, pp. 5282-5288, 2017.

\bibitem{Bech18}
C. P. Bechlioulis, M. A. Demetriou and K. J. Kyriakopoulos, ``A distributed control and parameter estimation protocol with prescribed performance for homogeneous lagrangian multi-agent systems", \emph{Autonomous Robots}, pp. 1-17, 2018.

\bibitem{Ver18}
C. K. Verginis, C. P. Bechlioulis, D. V. Dimarogonas and K. J. Kyriakopoulos, ``Robust distributed control protocols for large vehicular platoons with prescribed transient and steady-state performance", \emph{IEEE Trans. Control Syst. Technol.}, Vol. 26, pp. 299-304, 2018.

\bibitem{Gui17}
M. Guinaldo and D. V. Dimarogonas, ``A hybrid systems framework for multi agent task planning and control", \emph{in Proc. American Control Conference (ACC)}, pp. 1181-1186, 2017.

\bibitem{Lars17}
L. Lindemann, C. K. Verginis and D. V. Dimarogonas, ``Prescribed performance control for signal temporal logic specifications", \emph{in Proc. Decision and Control (CDC)}, pp. 2997-3002, 2017.
%

\bibitem{Abde98}
T. Abdelzaher and K. G. Shin, ``End-host architecture for QoS-adaptive communication", \emph{in Proc. Real-Time Technology and Applications Symposium}, pp. 121-130, 1998.

\bibitem{Lu02}
C. Lu, J. A. Stankovic, S. H. Son and G. Tao, ``Feedback control real-time scheduling: Framework, modeling, and algorithms", \emph{Real-Time Systems}, Vol. 23, pp. 85-126, 2002.

\bibitem{Meng16}
M. Guo, J. Tumova and D. V. Dimarogonas, ``Communication-free multi-agent control under local temporal tasks and relative-distance constraints", \emph{IEEE Trans. Autom. Control}, Vol. 61, pp. 3948-3962, 2016.

\bibitem{Zach05}
P. T. Zacharia and N. A. Aspragathos, ``Optimal robot task scheduling based on genetic algorithms", \emph{Rob. Comput. Integr. Manuf.}, Vol. 21, pp. 67-79, 2005.

\bibitem{Oma10}
F. A. Omara and M. M. Arafa, ``Genetic algorithms for task scheduling problem", \emph{J. Parallel Distrib. Comput.}, Vol. 70, pp. 13-22, 2010.

\bibitem{Mesb10}
M. Mesbahi and M. Egerstedt, ``Graph theoretic methods in multiagent networks", \emph{Princeton University Press}, 2010.

\end{thebibliography}
\end{document}